# Simultaneous ultraviolet, visible and near-infrared continuous-wave lasing in a rare-earth-doped microcavity

Bo Jiang, Song Zhu, Linhao Ren, Lei Shi,[*] and Xinliang Zhang

Wuhan National Laboratory for Optoelectronics, Huazhong University of Science and Technology, Wuhan 430074, China

[*]Corresponding Author: lshi@hust.edu.cn

**ABSTRACT:**

Microlaser with multiple lasing bands is critical in various applications, such as full-colour display, optical communications and computing. Here, we propose a simple and efficient method for homogeneously doping rare earth elements into a silica whispering-gallery-mode microcavity. By this method, we demonstrate simultaneous and stable lasing covering ultraviolet, visible and near-infrared bands in an ultrahigh-Q (exceeding $10^8$) Er-Yb co-doped silica microsphere under room temperature and continuous-wave pump for the first time. The lasing thresholds of the 380, 410, 450, 560, 660, 800, 1080 and 1550 nm-bands are estimated to be 380, 150, 2.5, 12, 0.17, 1.7, 10 and 38 μW, respectively, where the lasing in the 380, 410 and 450 nm-bands by Er element have not been separately demonstrated under room temperature and continuous-wave pump until this work. This ultrahigh-Q doped microcavity is an excellent platform for high-performance multi-band microlasers, ultrahigh-precise sensors, optical memories and cavity-enhanced light-matter interaction studies.

**INTRODUCTION**

Multi-band lasers are attractive for their potential applications in color display, lighting, wavelength-division multiplexing, optical communications and computing[1-4]. Rare earth (RE) elements with abundant long-lived intermediate energy levels and intraconfigurational transitions, can emit light emission from deep-ultraviolet to mid-infrared by pumping high-energy photons or low-energy photons, i.e., downshifting and upconversion, respectively, and thus can be applied in data storage[5-7], three-dimensional display[8,9], luminescent biomarkers[10-12] and lasers[13-15]. Whispering-gallery-mode (WGM) optical microcavity with ultrahigh quality (Q) factor and small mode volume exhibiting ultrahigh energy density and strong light-matter interaction, can be applied in sensing[16-18], nonlinear optics[19-24], cavity optomechanics[25,26] and lasers[27-33]. Therefore, doping RE elements into an ultrahigh-Q microcavity without degrading its intrinsic Q factor is an effective way to construct a low-threshold and narrow-linewidth multi-band microlaser[29,30].

By implanting erbium (Er) ions into a microdisk cavity under 2 MeV acceleration, Kippenberg et al. have realized a microlaser in the 1550 nm-band with the threshold of 43 μW [32] for resonant pump. Yang et al. processed a sol-gel based Er-silica film to form a doped microtoroid cavity, achieving the lowest threshold of 660 nW in the 1550 nm-band[33] under resonant pump. Besides the luminescence in the 1550 nm-band, the upconversion luminescence by Er ions also attracts great interest. Klitzing et al. have achieved green upconversion lasing with the threshold of 30 μW under resonant pump by an Er-fluoride-glass microsphere[34]. Based on the sol-gel technology, Lu et al. fabricated a highly doped Er microtoroid cavity with Q factor of $10^7$, realizing green upconversion laser with the threshold as low as 690 μW for resonant pump[35]. Recently, Bravo et al. have achieved CW red laser from an Er-upconverting-nanoparticle-coated plasmon cavity with a threshold of 70 W/cm$^2$.[36] Zhu et al. employed NaYF$_4$:Yb/Er@NaYF$_4$ core-shell upconversion nanocrystals as gain material, realizing simultaneous lasing in the 410, 540 and 655 nm-bands by a bottle-like WGM microcavity under a 3-pulse pump system.[37] Spontaneous upconversion emission of Er ions has been widely investigated from deep-ultraviolet to near-infrared. However, to the best of our knowledge, under room-temperature and CW pump, stimulated ultraviolet, violet and blue upconversion emissions have not been separately reported. The main reason should be that the inhomogeneity of the doped cavity will cause extremely large scattering loss in the short-wavelength band. Therefore, a doped cavity with ultrahigh Q factor should be the key to achieve the short-wavelength lasing.

In this work, we demonstrate a polymer assisted method for doping RE ions homogeneously. As a consequence, an ultrahigh-Q ($Q_{intrinsic}$=1.34×10$^8$ in the 1550 nm-band) Er-Yb co-doped silica microsphere cavity is fabricated, achieving simultaneous CW lasing in ultraviolet, visible and near-infrared bands at room temperature for the first time. The thresholds under nonresonant pump are estimated to be 380, 150, 2.5, 12, 0.17, 1.7, 10 and 38 μW for the 380, 410, 450, 560, 660, 800, 1080

and 1550 nm-bands, respectively, of which, the 380, 410 and 450 nm-bands CW lasing by Er ions have not been separately demonstrated under room temperature pump until this work. The Er-Yb co-doped microsphere also shows great lasing stability over 190 minutes, indicating a stable and high-performance multi-band laser source for practical applications. Besides, this active microcavity with ultrahigh Q factor is also an extraordinary platform for ultrahigh-precise sensing[38,39], optical memories[40], and cavity-enhanced light-matter interaction studies.

**Device characterization and experiment setup**

The details of the design and fabrication of an Er-Yb co-doped microsphere cavity are described in the Supplementary Section I. As shown in Fig. 1a, the Er-Yb co-doped microsphere cavity with a diameter of about 57 μm is coupled with an optical microfiber with a diameter of about 1.2 μm. The microsphere cavity is pumped by a 975-nm CW semiconductor laser with a linewidth of 1 nm, producing simultaneous ultraviolet, visible and near-infrared CW lasing. In this pump scheme, the pump laser with a linewidth of 1 nm can cover several resonant wavelengths. Only a small fraction of the pump power can be coupled into the cavity, thus this pump scheme is called nonresonant pump. In contrast, for resonant pump, most of the pump power can be coupled into the resonance mode of the cavity and thus significantly enhanced, achieving extremely high energy density for low lasing threshold and high pump efficiency[29]. However, the resonant pump requires a wavelength-fine-tunable narrow-linewidth laser source. The inset of Fig. 1a is a charge coupled device (CCD) image of the microlaser with a pump power of 4 mW, in which green light luminescence is dominant while red light is glimmering (the response of our CCD at 665 nm is weak). The layers of the bright rings can be recognized as multiple high-order WGMs. The homogeneous bright rings distribute without obvious scattering bright spots induced by the ion clusters, indicating the homogeneous RE ion doping. White-like light can be seen in the tail of the microsphere, which should be attributed to the blue, green and red leaky modes.

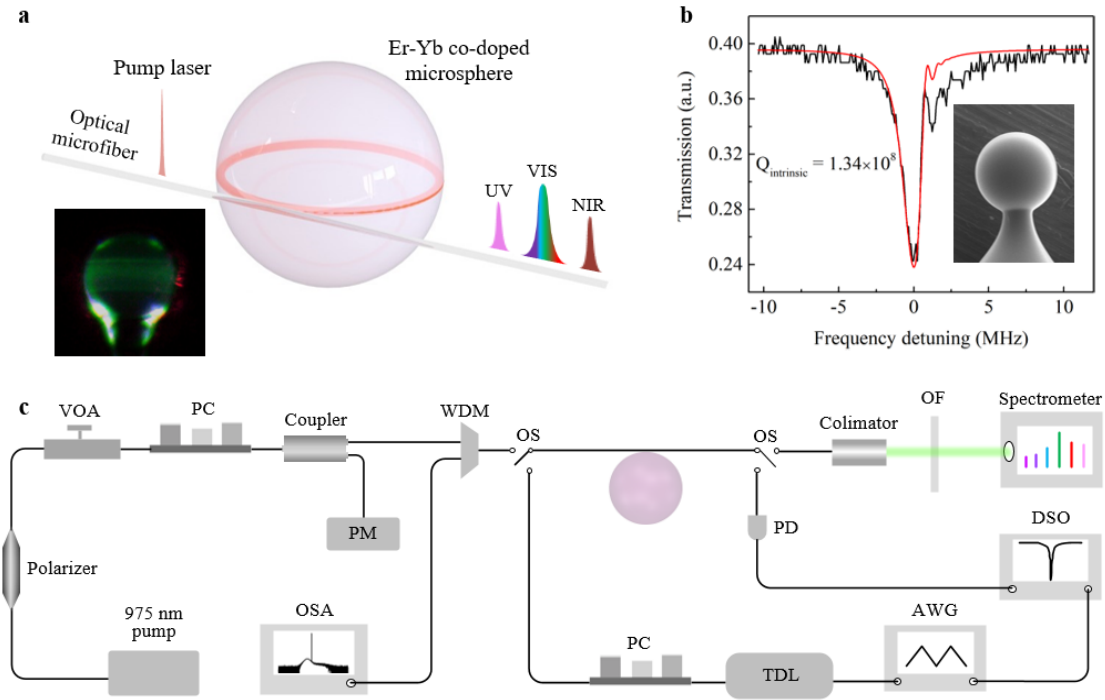

**Fig. 1 | Simultaneous ultraviolet, visible and near-infrared CW lasing at room temperature. a**, Schematic of ultraviolet, visible and near-infrared lasing by an Er-Yb co-doped microsphere under 975-nm CW laser excitation. Inset: an optical image of the Er-Yb co-doped microsphere with a pump power of 4 mW. **b**, Q factor characterization of the device in the 1550 nm-band. The black and red lines represent the experimental measurement and the theoretical fitting, respectively. Inset: a scanning electron microscopy image of the microsphere. **c,** Experimental set-up. VOA, variable optical attenuator; PM, power meter; PC, polarization controller; WDM, wavelength division multiplexer; OSA, optical spectrum analyser; OS, optical switch; OF, 975-nm band-stop optical filter; DSO, digital storage oscilloscope; TDL, tunable diode laser; AWG, arbitrary waveform generator; PD, photodetector.

High Q factor of the microcavity plays an essential role for achieving short-wavelength upconversion laser. The intrinsic Q factor of the doped silica microcavity is mainly limited by scattering loss, while scattering loss will extremely increase when the emission wavelength gets shorter. Therefore, a homogeneous RE ion doping method without introducing obvious ion clusters and defects is necessary. Fig. 1b shows the transmission spectrum operating around 1550 nm, in which the cavity ringdowm phenomenon can be clearly observed, which could occur in ultrahigh-Q cavities. The theoretical fitting indicates an ultrahigh intrinsic Q factor of $1.34\times10^8$, which is the highest Q factor among the Er-doped microcavities, and is very critical for achieving ultralow-threshold short-wavelength upconversion lasers.

All the experimental measurements were performed at room temperature. The structural symmetry of the microsphere cavity makes its eigenmodes two-fold degenerate. Two traveling-wave modes in the cavity propagate in opposite directions (clockwise and counterclockwise) with the same resonant frequency and polarization state. As shown in Fig. 1c, we measure the optical spectra of the clockwise light from 350 to 835 nm by a spectrometer. The counterclockwise light is coupled into an OSA via a WDM, and is used for the 1080 and 1550 nm-bands lasing spectrum acquisitions.

**Photoluminescence analysis of the Er-Yb co-doped silica microcavity**

A short-wavelength laser source can be realized by simultaneous two-photon and parametric oscillation processes. However, parametric oscillation requires phase match, and simultaneous two-photon absorption requires ultrahigh energy density. In contrast, the upconversion processes in RE elements can be realized with their long-lived intermediate energy levels, which does not demand phase match and ultrahigh energy density. As Er ions have abundant long-lived intermediate energy levels, manifold intraconfigurational transitions, and homogeneous energy levels distributions, they have the potential to realize ultraviolet upconversion laser with a 975-nm pump. In general, the upconversion processes in RE elements are mainly contributed by excited state absorption (ESA), energy transfer (ET), photon avalanche (PA) and cooperative energy transfer (CET) processes[41]. Yb ions are used as the sensitizer to transfer energy to Er ions for improving upconversion luminescence.

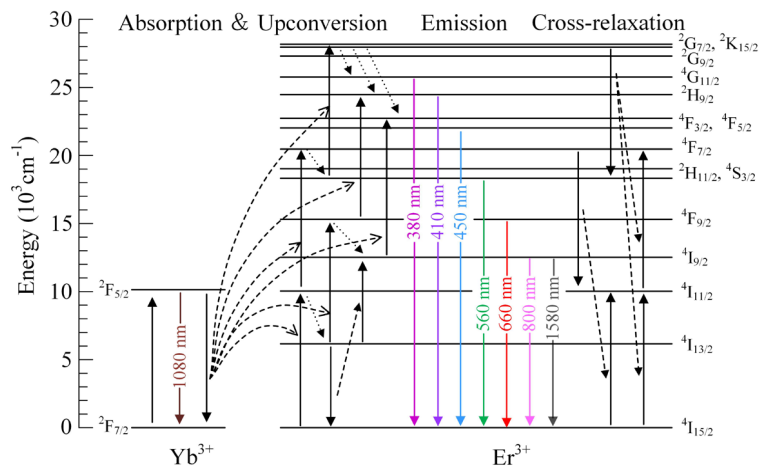

**Fig. 2 │ Energy level diagram and proposed light emission mechanisms.** The dashed and dotted arrows represent ET and multi-phonon relaxation processes, respectively. The upward full arrows represent the upward transition induced by photon excitation or ET process. The downward full arrows represent the downward transition induced by photon emission or ET process.

Fig. 2 shows the emission mechanisms of the 380 ($^4G_{11/2} \to {}^4I_{15/2}$, ultraviolet light), 410 ($^2H_{9/2} \to {}^4I_{15/2}$, violet light), 450 ($^4F_{3/2}, {}^4F_{5/2} \to {}^4I_{15/2}$, blue light), 560 ($^4S_{3/2} \to {}^4I_{15/2}$, green light), 660 ($^4F_{9/2} \to {}^4I_{15/2}$, red light), 800 ($^4I_{9/2} \to {}^4I_{15/2}$, near-infrared light), 1080 ($^2F_{3/2} \to {}^2F_{5/2}$, near-infrared light) and 1550 ($^4I_{13/2} \to {}^4I_{15/2}$, near-infrared light) nm-bands, in which the 1080-nm photons are emitted by Yb ions, while the others are emitted by Er ions. It is worth noting that both the $^2H_{11/2}$ and the $^4S_{3/2}$ states transit to the ground state will emit green photons. However, because these two energy levels have different lifetimes and populations, their lasing behaviors are different. In this work, the transition of $^4S_{3/2} \to {}^4I_{15/2}$ is the only consideration for green lasing. Under the 975-nm laser excitation, firstly, ground state absorption (GSA) induces the transitions of $^2F_{7/2} \xrightarrow{GSA} {}^2F_{5/2}$ (Yb ions) and $^4I_{15/2} \xrightarrow{GSA} {}^4I_{11/2}$ (Er ions). Subsequently, a portion of the $^4I_{11/2}$ states transit to the $^4I_{13/2}$ states rapidly with phonon-assisted non-radiative relaxation (NR). Then, ESA process in Er ions and ET process from the excited Yb and Er ions to the adjacent excited Er ions will resonantly generate upward transitions. The main upconversion processes are provided as follows: $^4I_{11/2} \xrightarrow{ESA/ET} {}^4F_{7/2} \xrightarrow{NR} {}^4S_{3/2} \xrightarrow{ESA/ET} {}^2G_{7/2}, {}^2K_{15/2} \xrightarrow{NR} {}^4G_{11/2}, {}^2H_{9/2}, {}^4F_{3/2}, {}^4F_{5/2}$, $^4I_{13/2} \xrightarrow{ET} {}^4I_{9/2} \xrightarrow{ESA/ET} {}^4F_{3/2}, {}^4F_{5/2}$ (the transition of $^4I_{13/2} \to {}^4I_{9/2}$ refers to ET between two Er ions of $^4I_{13/2}$) and $^4I_{13/2} \xrightarrow{ESA/ET} {}^4F_{9/2} \xrightarrow{EAS/ET} {}^2H_{9/2}$.

**Lasing characterization of the Er-Yb co-doped microcavity**

Considering that mode spacing is proportional to square of wavelength, the microsphere with a 57-μm diameter has a large number of resonance modes in short-wavelength band. In this situation, the competition among the lasing modes is unpredictable, and the lasing modes cannot be clearly distinguished by the spectrometer, thus we characterize the emission intensity as the integration of the whole lasing band instead of the intensity of a certain lasing mode. Fig. 3 shows the output intensity versus the pump power for the eight emission bands. At the low pump power, spontaneous emission intensity increases slowly at the beginning. When the pump power reaches the threshold, the slope of the curve will increase superlinearly, exhibiting an abrupt change, and indicating the lasing onset. Further increasing the pump power, the slope of the curve will decrease, as the emission intensity tends to be saturated. The whole curve exhibits a typical S-like behavior. The lasing thresholds are estimated through fitting the data dots by straight lines, and the crossover point of spontaneous emission and stimulated emission regions is defined as the threshold. Fig. 3a shows the thresholds of the 380 and 410 nm-bands, which are about 380 and 150 μW, respectively. Such low-threshold CW upconversion lasers under room-temperature indicate a feasible approach to obtain ultraviolet and violet lasers by using mature near-infrared semiconductor lasers. The thresholds of the 450 and 800 nm-bands are excited state absorption (ESA), energy transfer (ET), photon avalanche (PA) and excited state absorption (ESA), energy transfer (ET), photon avalanche (PA) and

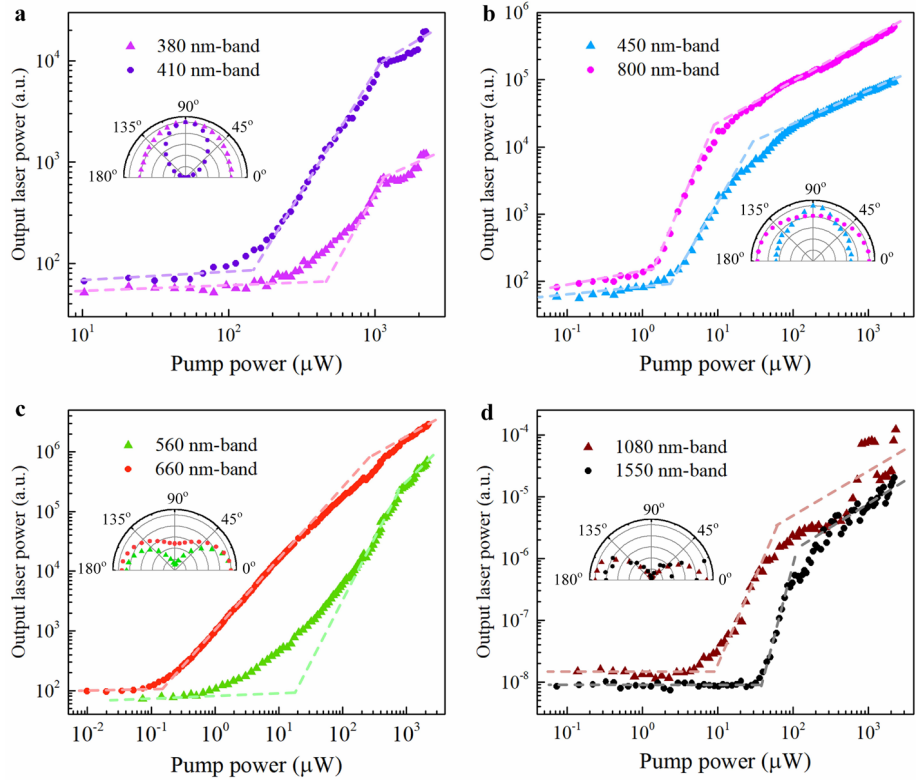

**Fig. 3 | Thresholds and polarization characteristics of ultraviolet, visible and near-infrared CW lasing.** The emission intensity versus the pump power in log-log scale. Inset: the emission intensity as a function of the polarization angel. The emission intensities are integrated from **a,** 380 to 396 nm and 402 to 426 nm, **b,** 450 to 489 nm and 756 to 816 nm, **c,** 555 to 575 nm and 645 to 590 nm, **d,** 1040 to 1090 nm and 1525 to 1590 nm.

$^4I_{9/2} \xrightarrow{ESA/ET} {}^4F_{3/2}, {}^4F_{5/2}$, their lasing curves are dependent and similar. Fig. 3c indicates that the lasing thresholds of the 560 and 660 nm-bands are 12 and 0.17 μW, respectively. The threshold of the 560 nm-band is lower than that obtained in the Er-doped silica microtoroid[35] and the Er-fluoride glass microsphere[34] under resonant pump. An extraordinarily low threshold of the 660 nm-band is estimated, which is about one-seventieth of the threshold of the 560 nm-band. This contrast is attributed to the difference between their intermediate energy level lifetimes during the

upconversion processes. The lifetime of the $^4I_{11/2}$ are much shorter than that of the $^4I_{15/2}$, which means that the population of the $^4S_{3/2}$ demands the higher pump power to be inverted. Fig. 3d indicates that the lasing thresholds of the 1080 (emitted by Yb ions) and 1550 nm-bands are about 10 and 38 μW, respectively. The threshold of the 1550 nm-band is lower than one-fiftieth of the previously reported lowest threshold under nonresonant pump[42-44], and is close to the thresholds under resonant pump[32,45-47]. In many cases, the thresholds of upconversion lasers are higher than the thresholds of downshifting lasers. However, in this case, the thresholds of the downshifting lasers are higher than the thresholds of the 450, 660 and 800 nm-bands. The lasing threshold is mainly determined by pump efficiency, gain material induced absorption loss, cavity intrinsic loss and coupling loss. In our scheme, the Q factor of the Er-Yb co-doped microsphere comes up to $1.34\times10^8$ in the 1550 nm-band, which means the ultralow gain material induced absorption loss and cavity intrinsic loss. Thus, pump efficiency and coupling loss should be primarily considered for the 1080 and 1550 nm-bands lasing. Here, the 1080 and 1550 nm-band lasing modes are over-coupled, resulting in high coupling loss and increases of the 1080 and 1550 nm-band lasing thresholds. In contrast, the upconversion lasing modes are located at the weak coupling region. However, the lifetimes of the upper states are much shorter than that of the $^4I_{15/2}$ states, as a result, in this ultrahigh-Q microcavity, inverting the populations to eliminate the gain material induced absorption loss should be primarily considered for the upconversion lasing. Therefore, the upconversion lasing thresholds are highly dependent on the numbers of the absorbed photons and the lifetimes of the intermediate energy levels during the upconversion processes.

For our experiment, around the lasing thresholds, the pump-output curves exhibit gradual slope changes. This should be attributed to large spontaneous emission coupling factors ($\beta$). Spontaneous emission coupling factor is defined as the ratio of the emission band coupled into the WGMs: $\beta = F_p/(1+F_p)$ [48], where $F_p$ represents the Purcell factor. And the Purcell factor can be expressed as[49]:

$$F_p = \frac{3}{4\pi^2} \frac{(\lambda_0/n_{eff})^3}{V_{eff}} Q_{eff}^{-1}$$

where $V_{eff}$ represents the effective mode volume, $n_{eff}$ represents the mode effective refractive index, $\lambda_0$ represents the resonant wavelength, and $Q_{eff}$ represents the quality factor of the system ($Q_{eff}^{-1} = Q_w^{-1} + Q_e^{-1}$, where $Q_w$ and $Q_e$ represent the quality factors of the WGM and the emitters, respectively). As $F_p$ is inversely proportional to $V_{eff}$, the enhanced spontaneous emission could occur in the microlaser.

The polarization state of the multi-band microlaser is also examined. The insets show that the variations of the relative emission intensities by rotating the polarizer from 0 degree to 180 degrees with a step size of 10 degrees. It can be found that the relative emission intensities of all lasing bands are symmetric with respect to 90 degrees, which indicates that they are polarized. Among them, the lasing lights of the 410, 560, 1080 and 1550 nm-bands are linearly polarized, and the lasing in the 380, 450, 660 and 800 nm-bands are elliptically polarized, which is attributed to that multiple lasing modes can be comprised by orthogonal polarized transverse electric (TE) and transverse magnetic (TM) modes.

To characterize the lasers more detailed, the lasing spectra under different pump powers are shown in Fig. 4. Under the 3-μW pump power excitation, the 660 and 800 nm-bands have reached the lasing thresholds, as shown in Fig. 4a. The peak around 488 nm is attributed to the transition of the $^4F_{7/2} \rightarrow {}^4I_{15/2}$. A portion of the states of the $^4F_{7/2}$ will transit to the $^2H_{11/2}$ and the $^4S_{3/2}$ states, and then emit green photons, as shown in Fig. 2. Under the 36-μW pump power excitation, the 450, 560 and 1080 nm-band lasers are achieved, as shown in Fig. 4b. Interestingly, there are also several wide linewidth peaks around 1535 nm, as shown in the inset of Fig. 4b. However, according to Fig. 3d, with this pump power below the lasing threshold, the 1550 nm-band lasing has not been achieved. These wide-linewidth peaks are attributed to the Purcell-enhanced spontaneous emission. With the Purcell effect, spontaneous emission will be enhanced, emerging some spontaneous emission peaks locating at the multiple adjacent resonance wavelengths of the cavity. However, the spacings of the adjacent peaks are too small to be resolved by the spectrometer with a resolution of 0.06 nm, eventually, some wide-linewidth peaks are exhibited. These spontaneous emission peaks can be easily distinguished with the laser peaks by their linewidths, which are about 0.8 and 0.06 nm respectively. As shown in Fig. 4c, under the pump power of 100 μW, the lasing in the 1550 nm-bands has been achieved. Er ions have two gain peaks around 1535 and 1550 nm ($^4I_{13/2} \rightarrow {}^4I_{15/2}$). However, in this Er-ion highly-doped scheme, under a weak pump, reabsorption effect makes the device lase at 1564.56 nm firstly. With the higher pump power, the reabsorption effect gets weak, and the laser peaks shift to the shorter wavelengths. As the pump power is increased to 1100 μW, the lasing in the 380 and 410 nm-bands are also achieved, as shown in Fig.

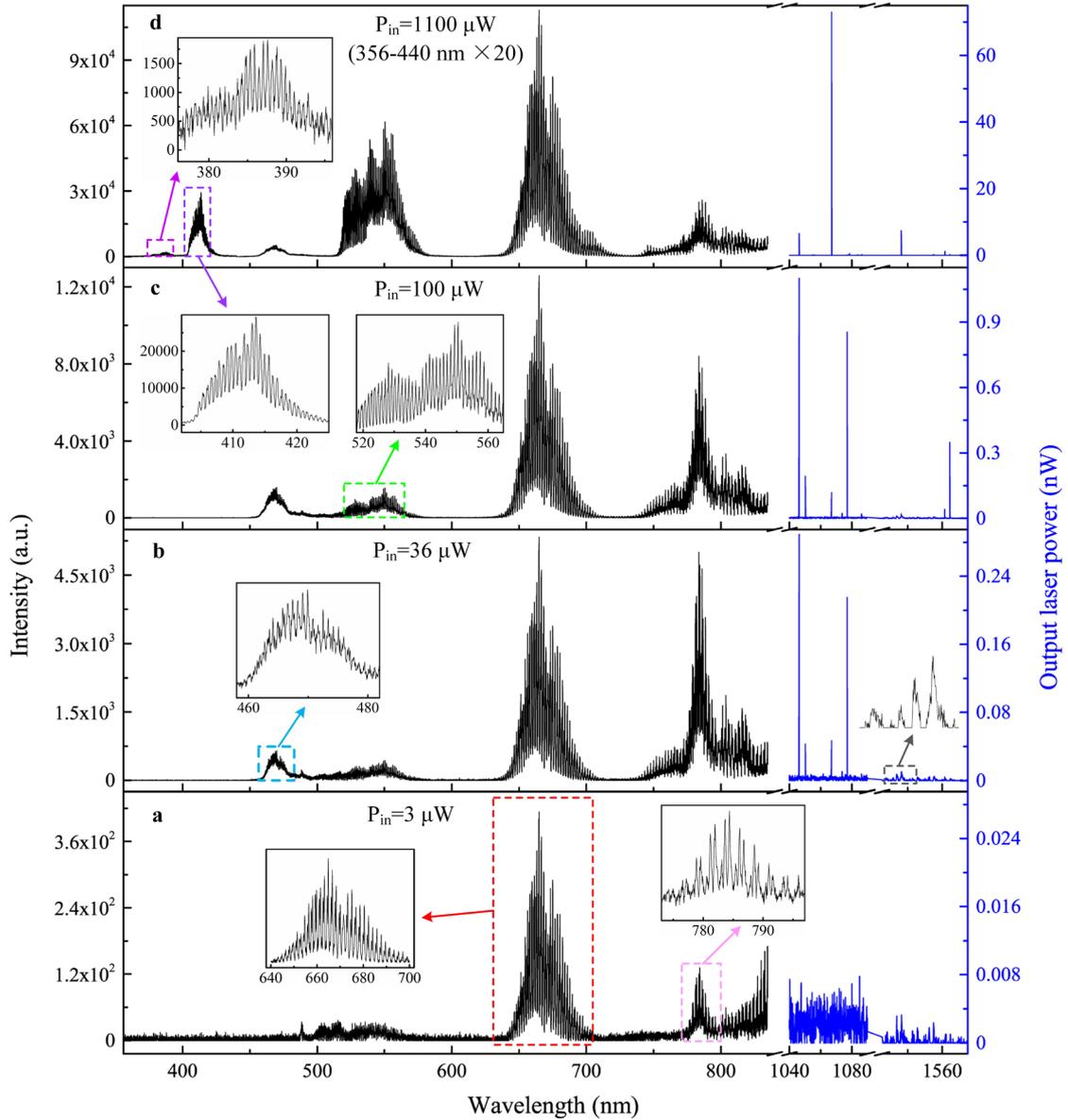

**Fig. 4 | Lasing spectrum evolution with the increased pump power.** The lasing spectra (black line) from 350 to 835 nm are acquired by the spectrometer. The lasing spectra (blue line) from 1040 to 1090 nm and 1525 to 1590 nm are acquired by the OSA. The pump powers are **a,** 3, **b,** 36, **c,** 100, **d,** 1100 μW, respectively, from the bottom to the top. The insets are the zoom-in lasing spectra.

4d. The laser intensities of the 380 and 410 nm-bands are relatively weak, compared with the other lasing bands. Except for the weaker intrinsic emission intensities, this is also due to the lower grating diffraction efficiencies and weaker coupling strengths between the microsphere and the microfiber in the 380 and 410 nm-bands, leading to that a small number of the emitted photons are coupled out and detected by the spectrometer (corresponding theoretical analysis is shown in Supplementary Section II). With the pump power increased from 100 to 1100 μW, the laser intensities of the 450, 800 and 1550 nm-bands increase slightly, while the laser intensities of the 560 and 660 nm-bands increase significantly. Under the increased pump power, a plenty of the $^4I_{15/2}$ and the $^4I_{13/2}$ states transit to the upper energy levels, as the result, the densities of the $^4I_{9/2}$, the $^4F_{3/2}$ and the $^4F_{5/2}$ states are restrained, as they are positively related to the density of the $^4I_{15/2}$ state. The lasing mode spacings of the 380, 410, 450, 560, 660, 800, 1080 and 1550 nm-bands are about 0.56, 0.642, 0.8, 1.19, 1.70, 2.44, 4.6 and 9.6 nm, which are well matched with the calculated free spectral ranges (FSRs) of 0.56, 0.65, 0.82, 1.22, 1.72, 2.53, 4.72 and 9.83 nm, respectively. The FSR is calculated by $FSR = \lambda_0^2 / \pi D n_g$, where $\lambda_0$ represents the center wavelength of the WGMs, $D$ represents the diameter of the cavity, $n_g$ represents the group refractive index.

One of the characteristics of the lasing onsets is distinct spectral linewidth narrowing around the lasing threshold. However, in our scheme, the laser linewidths of laser and Purcell enhanced spontaneous emission peaks are too narrow to be resolved by the spectrometer. Despite this fact, the mode-competition-induced evolutions of the lasing spectra can still be used to confirm the lasing onsets (the detail sees the Supplementary Section III).

Compared with downshifting lasing, upconversion lasing usually demands higher energy density. However, high pump intensity may cause thermal and optical damage for gain material, which leads to the failure of the emission intensity and the lasing wavelength shifting. Previous works about RE-based upconversion lasers usually employed pulsed laser pump or cryogenic environment to avoid thermal damage. For our CW lasing under room-temperature, we test their stabilities by measuring the variations of the lasing intensities and the lasing wavelengths over 190 minutes. As shown in Fig. 5a, the lasing intensities of the 380 and 410 nm-bands did not degrade obviously, but with the variations less than 7.3% and 14.4% ((maximum-minimum)/initial value), respectively, exhibiting an impressive intensity stability. The lasing intensities of the 450, 800 and 1080 nm-bands degraded at the beginning, and then went stabilized. In contrast, the lasing intensities of the 560, 660 and 1550nm-bands increased slightly at the beginning, and also then went stabilized. As shown in the Fig. 5b, the maximum wavelength shifts for the 450, 600 and 800 nm-bands are 0.452, 0.47 and 0.9 nm, respectively, which corresponds to one or two pixel sizes of the spectrometer. This result indicates the good lasing intensity stability and an excellent lasing wavelength stability of the proposed microlaser, making it possible for practical applications.

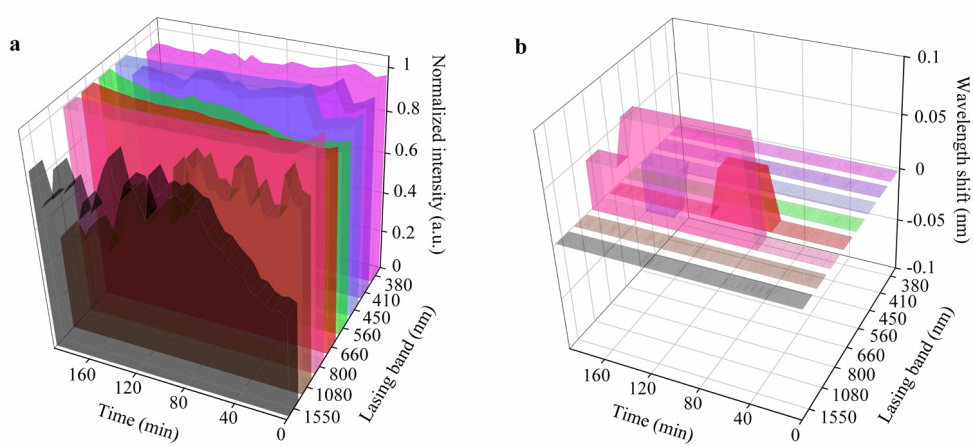

**Fig. 5 | Lasing stability. a.b.** Lasing intensity (a) and wavelength (b) evolutions over the duration of 190 minutes, exhibiting relatively stable emission intensities and slight wavelengths shifts.

**Conclusion**

In summary, we have proposed and demonstrated a simple and effective method for homogeneously doping RE elements into a silica microcavity. Through this method, we fabricated a 57-μm-diameter Er-Yb co-doped silica microsphere with the highest Q factor among the RE-doped microcavities up to $1.34\times10^8$ in the 1550 nm-band, achieving the simultaneous ultraviolet, visible and near-infrared CW lasing under room-temperature with the ultralow thresholds of 380, 150, 2.5, 12, 0.17, 1.7, 10 and 38 μW for the 380, 410, 450, 560, 660, 800, 1080 and 1550 nm-bands under nonresonant pump, respectively. This multi-band microlaser exhibits good emission intensity stability and excellent wavelength stability over 190 minutes. The significance of this demonstration can be understood in several aspects. Firstly, simultaneous ultraviolet, visible and near-infrared lasing have been achieved by an individual microcavity, of which lasing over a wavelength range of 1170 nm by a single gain material (Er element) has been realized. Among them, the stable upconversion lasing in the 380, 410 and 450 nm-bands are firstly exhibited under room-temperature and CW pump. Secondly, benefited from the ultrahigh Q factor of the doped microcavity, the multi-band microlaser has narrow linewidths and ultralow lasing thresholds, exhibiting extraordinary lasing performances. Thirdly, the desired lasers can be obtained by different RE elements via upconversion or downshifting processes, which allows for flexible pump schemes and abundant lasing wavelengths. Beside these advantages in laser applications, this ultrahigh-Q doped microcavity is also an excellent platform for ultrahigh-precise sensing, optical memories and cavity-matter-light interaction investigations.

## Acknowledgments

This work was supported by the National Natural Science Foundation of China (91850115, 11774110), the Fundamental Research Funds for the Central Universities (HUST: 2019kfyXKJC036, 2019kfyRCPY092), the State Key Laboratory of Advanced Optical Communication Systems and Networks (SJTU) (2021GZKF003), and the Key Research and Development Program of Hubei Province (2020BAA011).

**Methods**

**Device fabrication.** The fabrication details of the proposed microlasers are shown in Fig. S1.

**Device characterization.** The frequency detuning measurement is performed using a tunable diode laser (New Focus, TLB-6730-P) sweeping over the resonant modes. The laser wavelength scans around 1550 nm with a span of 22 GHz and a frequency of 50 Hz. To reduce the thermal nonlinear effect of the microcavity, the pump power is below 100 μW. The SEM image is performed by a Nova NanoSEM 450 (FEI Company). The CCD (Apico) image is captured using a 0.28 numerical aperture ×4 objective.

**Optical measurement.** The experiment setup is shown in Fig. 1. A 975-nm CW semiconductor laser (Fby photoelectric technology, OS974-500-HI1060-F) with a linewidth of about 1 nm was used to pump the Er-Yb co-doped microsphere via an optical microfiber. The counterclockwise 1080 and 1550 nm-bands lasers are coupled into an OSA (Anritsu, MS9740A) via a WDM (980/1550 nm). The clockwise upconversion lasers are coupled to an electric cooled silicon array charge-coupled device (Horiba, 1024X256-OE) equipped spectrometer (Horiba, iHR-550) by a collimator (Thorlab, F240FC-532), where a gating of 900 groove/mm blazed at 850 nm and a short-pass filter (Grand unified optics) are applied. As the width of the slit is set to 200 μm, the measurement resolution is estimated to be 0.33 nm (tested by a He-Ne 633 nm laser ~316 MHz, Newport, R-30993).

**Data availability.** The data that supports the findings of this study is available from the corresponding author on reasonable request.